\documentclass[prd,nofootinbib]{revtex4}
\usepackage{graphicx, epsfig}
\usepackage{color}
\usepackage{mathrsfs}
\usepackage{amsmath}

\newcommand{\be}{\begin{equation}}
\newcommand{\ee}{\end{equation}}
\newcommand{\bd}{\begin{equation*}}
\newcommand{\ed}{\end{equation*}}
\newcommand{\bea}{\begin{eqnarray}}
\newcommand{\eea}{\end{eqnarray}}

\newcommand{\gapp}{\mathrel{\raise.3ex\hbox{$>$}\mkern-14mu
              \lower0.6ex\hbox{$\sim$}}}
\newcommand{\lapp}{\mathrel{\raise.3ex\hbox{$<$}\mkern-14mu
              \lower0.6ex\hbox{$\sim$}}}

\begin{document}

\title{Classical and Quantum Equations of Motion of an $n$-dimesional BTZ Black Hole}
\author{Eric Greenwood}
\affiliation{Department of Geology and Physics, University of Southern Indiana, Evansville, IN 47712}
\begin{abstract}

We investigate the gravitational collapse of a non-rotating $n$-dimensional BTZ black hole in AdS 
space in the context of both classical and quantum mechanics. This is done by first deriving the 
conserved mass of a ``spherically" symmetric domain wall, which is taken as the classical 
Hamiltonian of the black hole. Upon deriving the conserved mass, we also point out that, for a 
``spherically" symmetric shell, there is an easy and straight-forward way of determining the 
conserved mass, which is related to the proper time derivative of the interior and exterior times. 
This method for determining the conserved mass is generic to any situation (i.e.~any equation of 
state), since it only depends on the energy per unit area, $\sigma$, of the shell.

Classically, we show that the time taken for gravitational collapse follows that of the typical 
formation of a black hole via gravitational collapse; that is, an asymptotic observer will see that the 
collapse takes an infinite amount of time to occur, while an infalling observer will see the collapse 
to both the horizon and the classical singularity occur in a finite amount of time. 
Quantum mechanically, we take primary interest in the behavior of the collapse near the horizon and 
near the classical singularity from the point of view of both asymptotic and infalling observers. In 
the absence of radiation and fluctuations of the metric, quantum effects near the horizon do not 
change the classical conclusions for an asymptotic observer. The most interesting quantum 
mechanical effect comes in when investigating near the classical singularity. Here, we find, that 
the quantum effects in this region are able to remove the classical singularity at the origin, since the 
wave function is non-singular, and is also displays non-local effects, which depend on the energy 
density of the domain wall. 

\end{abstract}

\maketitle

\section{Introduction}

The question of gravitational collapse is always an interesting question in theoretical physics; 
whether it be to study the classical formation of a black hole 
\cite{Oppenheimer:1939ue,Vachaspati:2006ki,Greenwood:2008ht,Greenwood:2009gp}, quantum 
formation of a black hole \cite{Saini:2014qpa}, induced quasi-particle production 
\cite{Greenwood:2008zg}, or thermalization processes \cite{Danielsson:1999zt,Danielsson:1999fa} 
and different kinds of entropies \cite{Baron:2013cya,Baron:2012fv} within the context of the 
AdS/CFT correspondence. Moreover, due to the applications using the AdS/CFT correspondence, 
gravitational collapse in AdS has become of greater importance. Therefore, it is worth investigating 
the gravitational collapse of an $n$-dimensional, massive, BTZ black hole in AdS. In this paper 
we will investigate the gravitational collapse in the context of the Gauss-Codazzi equations. 

In Section \ref{sec:EQ} we review the Gauss-Codazzi equations. Here, we are interested in the 
Gauss-Codazzi equations for a general, spherically symmetric, $(n-1)$-dimensional hypersurface 
embedded in an $n$-dimensional space-time. Upon reviewing the Gauss-Codazzi equations, we 
also point out a straight-forward method for determining the conserved mass of the collapsing shell, 
which only depends on the proper-time derivative of the interior and external time coordinates. 
In Section \ref{sec:BTZ} we specialize to an $n$-dimensional BTZ black hole and derive the 
conserved mass using the straight-forward method found in Section \ref{sec:EQ}. Since the mass 
is a conserved quantity, we may treat it as the Hamiltonian of the collapsing shell, which we will 
use to derive the classical and quantum equations of motion for different observers. Here, 
we have chosen the most relevant observers; the asymptotic, which is stationary at spatial infinity, 
and infalling (one who is falling together with the collapsing domain wall). The quantum 
collapse for both observers are obtained by utilizing a minisuperspace version of the functional 
Schr\"odinger equation originally developed in Ref.~\cite{Vachaspati:2006ki}. In Section 
\ref{sec:Asymptotic} we study the classical and quantum equations of motion from the point of 
view of an asymptotic and in Section \ref{sec:Infalling}, we study the classical and quantum 
equations of motion from the viewpoint of an infalling observer.  As far as the quantum collapse is 
concerned, the most important domains of interest are the near horizon regime, for the asymptotic 
observer, and the near classical singularity regime, for the infalling observer. In Section 
\ref{sec:QuantumAsymptotic} we explore quantum effects in the near-horizon limit for an asymptotic 
observer and show that the horizon takes an infinite amount of time to form, in agreement with 
the classical result. In Section \ref{sec:QuantumInfalling} we explore the quantum effects near the 
classical singularity and demonstrate that the wavefunction describing the collapsing domain wall is 
non-singular at the origin with non-local effects, which were absent at large distances. Our results, 
however, are in the absence of quantum radiation and the fluctuations of the metric which may 
alter our results. 

\section{The Equations}
\label{sec:EQ}

In this section we will summarize the setup in \cite{Ipser:1983db} and follow the notation found 
therein. As in \cite{Ipser:1983db}, we will let $S$ denote an infinitely thin, and for our purposes, 
$(n-1)$-dimensional time-like hypersurface, which contains a delta function stress-energy, and let 
$\xi^a$ be its unit space-like normal ($\xi_a\xi^a=1$). The $(n-1)$-metric intrinsic to the hypersurface 
$S$ is 
\be
  h_{ab}=g_{ab}-\xi_a\xi_b
  \label{h}
\ee
where $g_{ab}$ is the $n$-metric of the space-time. Additionally, we take $\nabla_a$ to denote the 
covariant derivative associated with $g_{ab}$ and 
\be
  D_a=h_a{}^b\nabla_b
  \label{D_a}
\ee
be the covariant derivative on the induced $(n-1)$-dimensional hypersurface. The extrinsic 
curvature of $S$, denoted by $\Pi_{ab}$, is defined by
\be
  \Pi_{ab}\equiv D_a\xi_b=\Pi_{ba}.
  \label{pi}
\ee

The contracted forms of the first and second Gauss-Codazzi equations are then given by
\bea
  ^{(n-1)}R+\Pi_{ab}\Pi^{ab}-\Pi^2&=&-2G_{ab}\xi^a\xi^b\label{Gauss}\\
  h_{ab}D_c\Pi^{ab}-D_a\Pi&=&G_{bc}h^b{}_a\xi^c\label{Codazzi}.
\eea
Here $^{(n-1)}R$ is the Ricci scalar curvature of the $(n-1)$-geometry, $\Pi$ is the trace of 
the extrinsic curvature, and $G_a{}^b$ is the Einstein tensor in $n$-dimensional space-time.

As mentioned above, we will be working with infinitely thin domain walls so that the stress-energy 
tensor $T_{ab}$ of the three-dimensional space-time has a $\delta$-function singularity across it. 
Since the extrinsic curvature is analogous to the gradient of the Newtonian gravitational potential, 
this in turn implies that the extrinsic curvature has a jump discontinuity across $S$. Therefore we can 
introduce the difference between the exterior and interior extrinsic curvatures as
\bd
  \Pi_{ab,+}-\Pi_{ab,-}=[\Pi_{ab}]
\ed
and the stress-energy on the hyper surface as the integral over the three-dimensional space-time 
stress-energy 
\bd
  S_{ab}\equiv\int dl\,T_{ab},
\ed
where $l$ is the proper distance through $S$ in the direction of the normal $\xi^a$, and where the subscripts $\pm$ refer to values just off the surface on the side determined by the direction of 
$\pm\xi^a$. Hence the direction for, say $+\xi^a$, will be in the direction of the exterior geometry of 
the domain wall. In general, these geometries will be different for the case of the spherically 
symmetric domain wall. Using Einstein's and the Gauss-Codazzi equations, one can show that 
(see \cite{Misner:1974qy})
\be
  S_{ab}=-\frac{1}{8\pi G_N}\left([\Pi_{ab}]-h_{ab}[\Pi_c{}^c]\right).
  \label{action}
\ee
We can also introduce the ``average" extrinsic curvature
\be
  \tilde{\Pi}_{ab}=\frac{1}{2}\left(\Pi_{ab,+}+\Pi_{ab,-}\right)
  \label{tilde_pi}
\ee
which will be important later.\footnote{Note that often in the literature, (\ref{tilde_pi}) is written as
\bd
  \tilde\Pi=\{\Pi\},
\ed
which is similar to the anti-commutating brackets in quantum mechanics.}


For our current purposes, we will restrict ourselves to sources for which the material source for the 
stress-energy tensor is that consisting of a perfect fluid, see \cite{Ipser:1983db},
\be
  S^{ab}=\sigma u^au^b-\eta\left(h^{ab}+u^au^b\right).
  \label{Stress-Energy}
\ee
In (\ref{Stress-Energy}), $u^a$ is the four-velocity of any observer whose world line lies within $S$ 
and who sees no energy flux in his local frame\footnote{The four-velocity $u^a$ is a time-like unit 
vector that is orthogonal to the space-like unit normal $\xi^a$, i.e.,
\be
  u_au^a=-1, \hspace{2mm} \xi_au^a=0, \hspace{2mm} \xi_a\xi^a=+1.
  \label{conditions}
\ee
}, 
$\sigma$ is the energy per unit area and $\eta$ is the tension measured by the observer. 
Typically, $\eta$ takes on one of two values\footnote{However, one is not restricted to only these 
two options, one can write down a general equation of state, for example $\eta=c\tau$, and consider 
additional scenarios.}; for a dust wall, it is well known that $\eta=0$, while for 
a domain wall $\eta=\sigma$. These solutions come from taking the difference of the Codazzi 
equation (\ref{Codazzi}) on opposite sides of $S$, and using (\ref{h}) and (\ref{conditions}), one finds
\bd
  h_{ac}D_bS^{cb}=0,
\ed
or using (\ref{Stress-Energy}), one acquires
\bd
  (\sigma-\eta)h_{ac}u^bD_bu^c+u_aD_b[(\sigma-\eta)u^b]-h_a{}^bD_b\eta=0.
\ed
However, contracting with $u^a$, we then obtain\footnote{This condition leads to the fact that the 
energy per unit area and the tension are equal in the domain wall case.}
\bd
  D_b(\sigma u^b)-\eta D_bu^b=0.
\ed
Hence, for dust we have
\bd
  D_b(\sigma u^b)=0,
\ed
which is just the rest mass of the collapsing shell. However, one can in addition to these, use different 
equations of state for the shell. 


Let the vector field $u^a$ be extended off $S$ in a smooth fashion. The acceleration is then
\bea
  a^b=u^a\nabla_au^b&=&(h^b{}_c+\xi^b\xi_c)u^a\nabla_au^c\nonumber\\
     &=&h^b{}_cu^a\nabla_au^c-\xi^bu^au^c\Pi_{ac},
\eea
which, by virtue of the extrinsic curvature, has a jump discontinuity across $S$. The perpendicular 
components of the accelerations of observers hovering just off $S$ on either side satisfy
\begin{align}
  \xi_bu^a\nabla_au^b\Big{|}_++\xi_bu^a\nabla_au^b\Big{|}_-=&-2u^au^b\tilde\Pi_{ab}\nonumber\\
     =&-\frac{2\eta}{\sigma}(h^{ab}+u^au^b)\tilde\Pi_{ab}-\frac{2}{\sigma}S^{ab}\tilde\Pi_{ab}
     \label{perp1}
\end{align}
and
\bea
  \xi_bu^a\nabla_au^b\Big{|}_+-\xi_bu^a\nabla_au^b\Big{|}_-&=&-u^au^b[\Pi_{ab}]\nonumber\\
     &=&4\pi(\sigma-2\eta).
     \label{perp2}
\eea
The second term on the right-hand side of (\ref{perp1}) takes into account the contributions to the 
energy-tensor $T_{ab}$ that are not confined to $S$, which are present in the vacuo on opposite 
sides of $S$. For example, if there is only mass present, then $T_{ab}$ vanishes off the shell, hence 
the second term is zero. However, in the case that charge is present in $S$, then $T_{ab}$ does 
not vanish in the exterior vacuum and the contribution to $T_{ab}$ outside can be taken from the 
Maxwell tensor. 

We now wish to determine the conserved mass of an $n$-dimensional, ``spherically" symmetric shell. 
Our initial calculation is general for any ``spherically" symmetric metric, however in the following 
sections we will specialize to considering AdS space-time for massive shells. 

For a ``spherical" shell of stress-energy, let the unit normal $\xi_+$ point in the outward radial direction. 
Here we take the interior and exterior metric as,
\bd
  (ds^2)_\pm=-f_\pm(r)dt_\pm^2+\frac{1}{f_\pm(r)}dr^2+r^2d\Omega_{n-2}
       \label{out_BTZ_metric}
\ed
The equation of the wall is, 
\bd
  r=R(t).
\ed
The components for $u^a$ and $\xi^a$ $(a=t_\pm,r$, and angular pieces in that order$)$ and thus 
given by
\be
  (u^a_\pm)=\left(\frac{\alpha_\pm}{f_\pm},R_{\tau},\vec 0_{n-2}\right), \,\,\, (\xi_a^\pm)=\left(-R_{\tau},\frac{\alpha_\pm}{f_\pm},\vec 0_{n-2}\right),
  \label{u_xi}
\ee
where $R_{\tau}\equiv dR/d\tau$ and $\tau$ is the proper time of an observer moving with four-velocity $u^a$ at the wall, $\vec 0_{n-2}$ is a $(n-2)$-dimensional zero vector, and 
\begin{align}
  \alpha_\pm\equiv&f_\pm t_{\pm,\tau}=\sqrt{f_\pm+R_{\tau}^2}.
  \label{alpha_BTZ}
\end{align}
These expressions and the definitions (\ref{D_a}), (\ref{pi}) and (\ref{tilde_pi}) imply that
\be
  (h^{ab}+u^au^b)\tilde{\Pi}_{ab}=(\xi^r{}_++\xi^r{}_-)\frac{1}{2R},
\ee
and
\be
  \xi_ba^b_\pm=\xi_bu^a\nabla_au^b\Big{|}_\pm=\frac{1}{\alpha_\pm}\left[R_{\tau\tau}+\frac{f_\pm'}{2}\right]
  \label{acceleration}
\ee
where
\bd
  f_\pm'=\frac{df_\pm(r)}{dr}\Big{|}_{r=R(\tau)}.
  \label{f_prime}
\ed
Substituting (\ref{acceleration}) into (\ref{perp1}) and (\ref{perp2}) then yields the equations of motion
\begin{align}
  (\alpha_++\alpha_-)R_{\tau\tau}=&-\frac{\eta}{\sigma}\frac{\alpha_+\alpha_-(\alpha_++\alpha_-)}{R}-\frac{\alpha_+ f_+'}{2}-\frac{\alpha_- f_-'}{2}-2\frac{\alpha_+\alpha_-}{\sigma}S^{ab}\tilde\Pi_{ab}\label{plus}\\
  (\alpha_--\alpha_+)R_{\tau\tau}=&4\pi\alpha_+\alpha_-(\sigma-2\eta)-\frac{\alpha_- f_-'}{2}+\frac{\alpha_+ f_+'}{2}.\label{minus}
\end{align}
Before we move on we will make some general comments on the structure of both (\ref{plus}) and 
(\ref{minus}). Here, we see that $R_{\tau\tau}$ is always negative provided $\eta,f_\pm'\geq0$. For 
our purposes, a circular domain wall with $\eta\geq0$ will always collapse to a black hole, regardless 
of its size. 

Taking the ratio of (\ref{plus}) and (\ref{minus}) allows us to eliminate $R_{\tau\tau}$ from the 
expression. After some algebra this yields
\be
  0=\sigma(\sigma-2\eta)+\frac{1}{4\pi(\alpha_++\alpha_-)}\left[\sigma(f_-'-f_+')+2\eta(f_--f_+)\right]+\frac{f_--f_+}{4\pi(\alpha_-+\alpha_+)^2}S^{ab}\tilde\Pi_{ab}.
  \label{sigma eq}
\ee

Here we note that (\ref{sigma eq}) may be factorized giving two roots which satisfy the equality, 
however only one of the two roots is valid. Alternatively, we can notice that from (\ref{action}) and 
(\ref{Stress-Energy}), we can write 
\bd
  -8\pi(\sigma-\eta)u_au_b+8\pi\eta h_{ab}=[\Pi_{ab}]-h_{ab}[\Pi_c{}^c],
\ed
or upon contracting with $u^au^b$, this reduces to
\be
  -8\pi\sigma=[\Pi_{ab}]u^au^b+[\Pi_c{}^c].
  \label{8pi}
\ee
Notice that the right-hand side of (\ref{8pi}) may be rewritten in terms of the extrinsic curvature 
\bd
  -8\pi\sigma=[\Pi_{ab}u^au^b]+[\Pi].
\ed
However, using (\ref{u_xi}), one can show that the trace of the extrinsic curvature takes the form
\bd
  \Pi_\pm=\frac{1}{\alpha_\pm}\left[R_{\tau\tau}+\frac{f_\pm'}{2}+(n-2)\frac{\alpha_\pm^2}{R}\right].
\ed
Thus, using (\ref{perp2}), we can then write (\ref{8pi}) as
\bd
  -8\pi\sigma=(n-2)\left(\frac{\alpha_+}{R}-\frac{\alpha_-}{R}\right)=(n-2)\frac{\alpha_+-\alpha_-}{R},
\ed
which finally yields
\be
  \alpha_--\alpha_+=\frac{8\pi\sigma R}{n-2}.
  \label{EOM}
\ee
Notice that (\ref{EOM}) only depends on the surface density of the collapsing shell, not on the 
tension. Thus, we can easily see that the solution for the collapsing shell is the same for both the 
dust shell and the domain wall, up to the spatial dependence of the surface density. Physically, there 
there is no distinction between these solutions, since for a domain wall $\sigma$ is a constant, while for 
a dust shell $\sigma R^{n-2}$ is a constant. Furthermore, one can show that (\ref{EOM}) yields the 
same result as the one valid root from (\ref{sigma eq}).

\section{BTZ Black Hole in $n$ Dimensions}
\label{sec:BTZ}

From \cite{Banados:1992wn}, we take that the exterior metric coefficient $f_+$ is given as
\be
  f_+=\frac{R^2}{\ell^2}-\frac{M}{R^{n-3}}
  \label{f+}
\ee
and the interior metric coefficient $f_-$ is given as
\be
  f_-=\frac{R^2}{\ell^2}.
  \label{f-}
\ee
From (\ref{EOM}), we can solve for the mass, which is then given by
\begin{align}
  M&=\frac{16\pi\sigma R^{n-2}}{n-2}\left(\alpha_--\frac{4\pi\sigma R}{n-2}\right)\nonumber\\
     &=\frac{16\pi\sigma R^{n-2}}{n-2}\left(\sqrt{f_-+R_\tau^2}-\frac{4\pi\sigma R}{n-2}\right)\nonumber\\
     &=\frac{16\pi\sigma R^{n-2}\sqrt{f_-}}{n-2}\left(\sqrt{1+\frac{R_\tau^2}{f_-}}-\frac{4\pi\sigma R}{(n-2)\sqrt{f_-}}\right)
  \label{BTZ_mass}
\end{align}
where in the last line we have rewritten the mass for later convenience. It can be checked that 
(\ref{BTZ_mass}) is a constant of motion, which is done in Appendix \ref{ch:check}.  

Since the mass in (\ref{BTZ_mass}) is conserved, the mass then represents the total energy of the 
collapsing shell and may be treated as the Hamiltonian of the system. We will then determine the 
equations of motion for the collapsing shell using the conserved mass in (\ref{BTZ_mass}) as the 
Hamiltonian system. However, since the Hamiltonian is not invariant under change of coordinate 
system, we will work with the Lagrangian for the system. The Lagrangian is given by,
\begin{align}
  L(\tau)&=-\frac{16\pi\sigma R^{n-2}}{n-2}\left(\alpha_--\frac{4\pi\sigma R}{n-2}-R_\tau\sinh^{-1}\sqrt{\frac{R_\tau^2}{f_-}}\right)\nonumber\\
     &=-\frac{16\pi\sigma R^{n-2}}{n-2}\left(\sqrt{f_-+R_\tau^2}-\frac{4\pi\sigma R}{n-2}-R_\tau\sinh^{-1}\sqrt{\frac{R_\tau^2}{f_-}}\right).
  \label{Lm}
\end{align}
\section{Asymptotic Observer}
\label{sec:Asymptotic}

In this section we will obtain the classical and quantum equations of motion as viewed by an 
asymptotic observer. 

Since we are interested in the asymptotic observer, we must transform the Lagrangian in (\ref{Lm}) 
to the coordinate time, which may be done by considering the effective action. Under the change of 
coordinates, the Lagrangian (\ref{Lm}) takes the form
\be
  L_m(t)=-\frac{16\pi\sigma R^{n-2}}{n-2}\left(\sqrt{\frac{f_-f_+^2}{\alpha_+^2}+\dot R^2}-\frac{4\pi\sigma Rf_+}{\alpha_+(n-2)}-\dot R\sinh^{-1}\left(\frac{\alpha_+}{f_+}\sqrt{\frac{\dot R^2}{f_-}}\right)\right)
  \label{Lm_t}
\ee
where $\dot R=dR/dt$ and $\alpha_+$ is given in (\ref{alpha_BTZ}), which can be rewritten in terms 
of $\dot R$
\be
  \frac{\alpha_+}{f_+}=\sqrt{\frac{f_+}{f_+^2-\dot R^2}}.
  \label{alpha_f}
\ee
Using (\ref{alpha_f}), we can rewrite (\ref{Lm}) as
\be
  L_m(t)=-\frac{16\pi\sigma R^{n-2}}{\sqrt{f_+}(n-2)}\left(\sqrt{f_-}\sqrt{f_+^2-(1-f_+/f_-)\dot R^2}-\frac{4\pi\sigma R}{n-2}\sqrt{f_+^2-\dot R^2}-\sqrt{f_+}\dot R\sinh^{-1}\left(\sqrt{\frac{f_+}{f_-}}\sqrt{\frac{\dot R^2}{f_+^2-\dot R^2}}\right)\right).
  \label{Lm_t}
\ee
The generalized momentum, $P$, may be derived from (\ref{Lm_t}), and is given by
\begin{align}
  P(t)=\frac{16\pi\sigma R^{n-2}}{\sqrt{f_+}(n-2)}\Big{(}&\sqrt{f_-}\frac{(1-f_+/f_-)\dot R}{\sqrt{f_+^2-(1-f_+/f_-)\dot R^2}}+\frac{f_+^2\dot R}{\sqrt{f_-}(f_+^3-\dot R^2)\sqrt{f_+^2-(1-f_+/f_-)\dot R^2}}\nonumber\\
     &+\sqrt{f_+}\sinh^{-1}\left(\sqrt{\frac{f_+}{f_-}}\sqrt{\frac{\dot R^2}{f_+^2-\dot R^2}}\right)-\frac{4\pi\sigma R\dot R}{(n-2)\sqrt{f_+^2-\dot R^2}}\Big{)}.
  \label{P_t}
\end{align}
Using (\ref{P_t}), we can then write the Hamiltonian in terms of $t$ as 
\be
  H(t)=\frac{16\pi\sigma R^{n-2}}{\sqrt{f_+}(n-2)}\left(\frac{\sqrt{f_-}f_+^2}{\sqrt{f_+^2-(1-f_+/f_-)\dot R^2}}+\frac{f_+^3\dot R^2}{\sqrt{f_-}(f_+^2-\dot R^2)\sqrt{f_+^2-(1-f_+/f_-)\dot R^2}}-\frac{4\pi\sigma Rf_+^2}{(n-2)\sqrt{f_+^2-\dot R^2}}\right).
  \label{H_t}
\ee

We are, however, interested in the near horizon evolution of the collapse; that is, the last moments 
of the collapse. Therefore, we are interested in when $R$ is close to $R_H$, where $R_H$ is the 
horizon radius, or when $f_+\to0$, in which case $f_-\to$ const, which in this case is 
$f_-=R_H^2/\ell^2$. In this limit, we can then rewrite (\ref{P_t}) as
\be
  P(t)\approx\frac{16\pi\mu R^{n-2}\dot R}{(n-2)\sqrt{f_+}\sqrt{f_+^2-\dot R^2}},
  \label{P_t approx}
\ee
where 
\be
  \mu\equiv\sigma\left(\sqrt{f_-}-\frac{4\pi\sigma R_H}{n-2}\right)
\ee
and $f_-$ is evaluated at $R=R_H$. In this limit, we may also rewrite the Hamiltonian (\ref{H_t}) as 
\be
  H(t)\approx\frac{16\pi\mu R^{n-2}f_+^{3/2}}{(n-2)\sqrt{f_+^2-\dot R^2}}.
  \label{H_t approx}
\ee
We may invert (\ref{P_t approx}) and solve for $\dot R$ as a function of $P(t)$ so that we may 
rewrite (\ref{H_t approx}) as
\be
  H=\sqrt{(f_+P)^2+\left(\frac{16\pi\mu R_H}{n-2}\right)^2}.
  \label{H_t P}
\ee

We are now in a position to determine the classical and quantum equations of motion for the collapse. 
Let us determine the classical equations of motion for the collapse first. 

\subsection{Classical Equations of Motion}

Since the Hamiltonian is a conserved quantity, from (\ref{H_t approx}) we can write 
\be
  \dot R=\pm f_+\sqrt{1-\frac{f_+R^{2(n-2)}}{h^2}},
\ee
where $h\equiv H(n-2)/16\pi\mu$, or in the near horizon limit becomes
\bd
  \dot R\approx\pm f_+\left(1-\frac{1}{2}\frac{f_+R^{2(n-2)}}{h^2}\right)\approx\pm f_+.
\ed
That is, the dynamics of the collapse are given by $\dot R\approx-f_+$, where the negative sign is 
chosen due to the fact that we are interested in collapse. Using (\ref{f+}), we then have
\bd
  R=R_H\tanh\left(\frac{R_Ht}{\ell^2}+\tanh^{-1}\frac{R_0}{R_H}\right),
\ed
where $R_0$ is the radius of the domain wall at $t=0$. We can see that, as far as the asymptotic 
observer is concerned, the classical solution implies that it takes an infinite amount of time for the 
horizon of the BTZ black hole to form, since $R(t)=R_H$ only as $t\to\infty$. 

\subsection{Quantum Equations of Motion}
\label{sec:QuantumAsymptotic}

From (\ref{H_t P}), we see that the quantum Hamiltonian, as far as the asymptotic observer is 
concerned, is the same as the quantum Hamiltonian found in 
\cite{Vachaspati:2006ki,Greenwood:2008zg,Greenwood:2009gp}, hence we can write the solution as 
a Gaussian wave-packet solution which is shrinking\footnote{Here, the Gaussian wave packet is 
shrinking due to the relationship between $R$ and $u$; that is, $dR=fdu$.} while it is propagating 
toward the horizon
\bd
  \Psi=\frac{1}{\sqrt{2\pi}s}e^{-(u+t)^2/2s^2},
\ed
where 
\bd
  u=\int\frac{dR}{f_+}
\ed
and $s$ is the width of the wave packet. In the $u$-coordinate, the horizon has been moved to 
$u=-\infty$, hence it takes an infinite amount of time for the wave packet to reach the horizon 
and hence does not contradict the classical equation of motion.

\section{Infalling Observer}
\label{sec:Infalling}

In this section we will obtain the classical and quantum equations of motion for both the massive and 
massive-charged shells as viewed by an infalling observer; that is, an observer who is riding along 
the shell. 

Starting with (\ref{Lm}) we may determine the generalized momentum for the infalling observer to 
be
\be
  P(\tau)=\frac{16\pi\sigma R^{n-2}}{n-2}\sinh^{-1}\sqrt{\frac{R_\tau^2}{f-}},
  \label{P_tau}
\ee
which may be inverted so that we may rewrite the Hamiltonian (\ref{BTZ_mass}) as
\be
  H(\tau)=\frac{16\pi\sigma R^{n-2}}{n-2}\left(\sqrt{f_-}\cosh\frac{P(\tau)}{16\pi\sigma R}-\frac{4\pi\sigma R}{n-2}\right).
  \label{H_tau P}
\ee

We may now determine the classical and quantum equations of motion for the collapse. As with the 
asymptotic observer, let's determine the classical equations of motion first.

\subsection{Classical Equations of Motion}

From (\ref{BTZ_mass}), the velocity of the domain wall, as far as the infalling observer is concerned, 
is given by 
\be
  R_\tau=\sqrt{\left(\frac{\tilde h}{R^{n-2}}+\frac{4\pi\sigma R}{n-2}\right)^2-f_-},
  \label{R_tau}
\ee
where $\tilde h=(n-2)H(\tau)/16\pi\sigma$. Even though the integral for $R$ may be performed exactly, 
the solution may not be inverted to solve for $R(\tau)$. Thus, we will seek an approximate solution. As 
a zeroth order approximation, (\ref{R_tau}) is just a constant in the region $R\sim R_H$, so that 
the solution is 
\bd
  R(\tau)=R_0-\tau\sqrt{\left(\frac{\tilde h}{R_H^{n-2}}+\frac{4\pi\sigma R_H}{n-2}\right)^2-\frac{R_H^2}{\ell^2}}
\ed
where again the minus sign is chosen due to the collapse of the shell and $R_0$ is again the initial 
position of the collapsing domain wall at $\tau=0$. As is expected, the infalling observer will see the 
horizon formed in a finite amount of time.

Another interesting limit is the time it takes for the domain wall to collapse down to the classical 
singularity, which is located at $R=0$. In the limit $R\to0$, the position of the domain wall goes 
as $R\approx\left(R_0^{n-1}-(n-1)\tilde h\tau\right)^{n-1}$, so that the time taken to reach the classical 
singular is also finite. 

\subsection{Quantum Equations of Motion}
\label{sec:QuantumInfalling}

For the infalling observer, the more interesting limit is the near singularity limit of the collapse, hence 
the $R\to0$ solution. From (\ref{R_tau}), near the classical singularity the velocity of the domain wall 
goes as
\bd
  R_\tau\sim-\frac{\tilde h}{R^{n-2}},
\ed
which is divergent. Since the velocity is negative near the classical singularity the Hamiltonian, 
written in terms of the conjugate momentum, may be written as
\be
  H=\frac{16\pi\sigma R^{n-2}}{n-2}\sqrt{f_-}\cosh\frac{P(\tau)}{16\pi\sigma R}=\frac{8\pi\sigma R^{n-2}}{n-2}\sqrt{f_-}e^{-\frac{P(\tau)}{16\pi\sigma R}}=\frac{8\pi\sigma R^{n-2}}{n-2}\sqrt{f_-}e^{\frac{i}{16\pi\sigma R}\frac{\partial}{\partial R}}.
  \label{H near}
\ee
We can see that (\ref{H near}) is non-local since it depends on an infinite number of derivatives, 
due to the $R^{-1}$ term in the exponential, and may not be truncated after a few derivatives, unlike 
a local Hamiltonian. 

Another way to see this is to note that (\ref{H near}) has the structure of a translation operator that 
generates an imaginary translation. If we define the new variable $z=R^2$, then we can rewrite 
(\ref{H near}) as
\be
  H=\frac{8\pi\sigma}{n-2}z^{n/2-1}\sqrt{f_-}e^{\frac{i}{8\pi\sigma}\frac{\partial}{\partial z}}.
  \label{H Near}
\ee
Hence (\ref{H Near}) translates wave function by a non-infinitesimal amount: $\frac{i}{8\pi\sigma}$ 
in $z$ and $\sqrt{\frac{i}{8\pi\sigma}}$ in $R$. That is, as the collapsing shell approaches the 
classical singularity, the wave function is related to its value at some distance point: 
$\Psi(R\to\sqrt{\frac{i}{8\pi\sigma}},\tau)$. 

At the classical singularity, on the other hand, (\ref{H Near}) simplifies to 
\bd
  \frac{\partial\Psi(R\to0,\tau)}{\partial\tau}=0,
\ed
where the wave function at some distant point is finite. From this, we can then see that the wave 
function is constant and finite at the origin. 


\section{Conclusion}
\label{sec:conclusion}

In this paper we studied the gravitational collapse of an $n$-dimensional, spherically symmetric, 
BTZ black hole, which is represented by an infinitely thin domain wall using the Gauss-Codazzi 
formalism. Interestingly, along the way we showed that, at least in the context of a spherically 
symmetric dust shell or domain wall, the equation of motion of the shell\footnote{In this paper we 
concentrated on the domain wall, however the formalism will work for either the dust shell, the 
domain wall or any equation of state. The overall difference is the spatial dependence of the 
energy density of the shell, which is constant for the domain wall and position dependent for the 
dust shell.} can be straightforwardly determined from the proper time derivatives of the interior and 
exterior time coordinates, see (\ref{EOM}). The Gauss-Codazzi formalism allowed us to determine the 
conserved mass of the domain wall, which may be interpreted as the Hamiltonian of the system. 
Using the conserved mass, we then determined the both the classical and quantum equations of 
motion for the domain wall in locations of interest from different viewpoints. As far as the collapse is 
concerned, the most relevant observers are the asymptotic observer and the infalling observer. 

In Section \ref{sec:Asymptotic} we studied the collapse from the viewpoint of an asymptotic observer, 
both classically and quantum mechanically. Classically, we found that the horizon is only formed after 
an infinite amount of observer time, independent of the number of dimensions, as is expected from the 
infinite gravitational redshift associated with the formation of the horizon. Quantum mechanically, we 
found that, in the absence of Hawking radiation and back reaction on the metric, again the horizon is 
only formed after an infinite amount of observer time. This result implies that simply quantizing the 
matter shell doesn't lead to fluctuations of the horizon which will allow the horizon to be formed in a 
finite amount of observer time.\footnote{This is, of course, not unexpected since we have not 
quantized the geometry and did not allow for back reaction of the metric, which would also lead to 
fluctuations of the horizon.} 

In Section \ref{sec:Infalling} we studied the collapse from the viewpoint of an infalling observer, 
which is an observer who is riding on the surface of the domain wall, hence is determined by the 
proper time of the shell. Classically, we found that the horizon is formed in a finite amount of proper time, regardless of dimensionality, which is expected since locally the metric is flat and the horizon 
does not present itself as a problem for this observer. Quantum mechanically, 
we investigated the collapse of the domain wall as it neared the classical singularity, $R\to0$. 
Here, we found that the Hamiltonian takes on the form of a translation operator, (\ref{H near}), 
which translates the wave function by a non-infinitesimal amount. That is to say, the wave 
function at a given point depends on the wave function at a distant point. The distant point depends 
on the energy density of the domain wall; that is, the smaller the energy density, the further the 
point is. Thus, the wave function displays non-local behavior.\footnote{Another way to see 
this is by investigating (\ref{H near}). Since the differential operator enters into the Hamiltonian 
via an exponential, one would usually Taylor expand the exponential and then truncate the 
series after a finite amount of terms. However, in (\ref{H near}), there is a $R^{-1}$ dependence, 
which means that as $R\to0$, the higher order terms become more important, instead of less 
important, meaning that would would have to keep all infinite number of terms to fully define 
the Taylor expansion. This is a sign of non-local behavior.} Furthermore, we found that the wave 
function is finite and constant near the classical singularity, regardless of the dimensionality. 



\appendix

\section{Check that Mass per unit Length is a constant of Motion}
\label{ch:check}

Here we wish to check that (\ref{BTZ_mass}) is a constant of motion. First, note that we 
can rewrite (\ref{BTZ_mass}) as
\bd
  M=\frac{16\pi\sigma R^{n-2}}{n-2}\left(\alpha_--\frac{4\pi\sigma R}{n-2}\right),
\ed
or using (\ref{EOM}) this becomes
\be
  M=R^{n-3}\left(\alpha_-^2-\alpha_+^2\right).
  \label{mass_alpha}
\ee
Taking the proper time derivative of (\ref{mass_alpha}) we obtain
\bd
  M_\tau=(n-3)R^{n-4}R_\tau\left(\alpha_-^2-\alpha_+^2\right)+2R^{n-3}\left(\alpha_-\alpha_{-,\tau}-\alpha_+\alpha_{+,\tau}\right).
\ed
Using (\ref{alpha_BTZ}) we have
\bd
  M_\tau=R_\tau\left[(n-3)R^{n-4}R_\tau\left(f_--f_+\right)+R^{n-3}\left(f_-'-f_+'\right)\right]
\ed
and using (\ref{f+}) and (\ref{f-}) we finally obtain
\bd
  M_\tau=0.
\ed
Since the proper time derivative is zero, this proves that the mass is a conserved quantity.


\begin{thebibliography}{99}

\bibitem{Oppenheimer:1939ue} 
  J.~R.~Oppenheimer and H.~Snyder,
  Phys.\ Rev.\  {\bf 56}, 455 (1939).

\bibitem{Vachaspati:2006ki} 
  T.~Vachaspati, D.~Stojkovic and L.~M.~Krauss,
  Phys.\ Rev.\ D {\bf 76}, 024005 (2007)
  [gr-qc/0609024].

\bibitem{Greenwood:2008ht} 
  E.~Greenwood and D.~Stojkovic,
  JHEP {\bf 0806}, 042 (2008)
  [arXiv:0802.4087 [gr-qc]].

\bibitem{Greenwood:2009gp} 
  E.~Greenwood, E.~Halstead and P.~Hao,
  JHEP {\bf 1002}, 044 (2010)
  [arXiv:0912.1860 [gr-qc]].

\bibitem{Saini:2014qpa} 
  A.~Saini and D.~Stojkovic,
  Phys.\ Rev.\ D {\bf 89}, no. 4, 044003 (2014)
  [arXiv:1401.6182 [gr-qc]].

\bibitem{Greenwood:2008zg} 
  E.~Greenwood and D.~Stojkovic,
  JHEP {\bf 0909}, 058 (2009)
  [arXiv:0806.0628 [gr-qc]].

\bibitem{Danielsson:1999zt} 
  U.~H.~Danielsson, E.~Keski-Vakkuri and M.~Kruczenski,
  Nucl.\ Phys.\ B {\bf 563}, 279 (1999)
  [hep-th/9905227].

\bibitem{Danielsson:1999fa} 
  U.~H.~Danielsson, E.~Keski-Vakkuri and M.~Kruczenski,
  JHEP {\bf 0002}, 039 (2000)
  [hep-th/9912209].

\bibitem{Baron:2013cya} 
  W.~H.~Baron and M.~Schvellinger,
  JHEP {\bf 1308}, 035 (2013)
  [arXiv:1305.2237 [hep-th]].

\bibitem{Baron:2012fv} 
  W.~Baron, D.~Galante and M.~Schvellinger,
  JHEP {\bf 1303}, 070 (2013)
  [arXiv:1212.5234 [hep-th]].

\bibitem{Ipser:1983db} 
  J.~Ipser and P.~Sikivie,
  Phys.\ Rev.\ D {\bf 30}, 712 (1984).

\bibitem{Misner:1974qy} 
  C.~W.~Misner, K.~S.~Thorne and J.~A.~Wheeler,
  San Francisco 1973, 1279p
  
\bibitem{Banados:1992wn} 
  M.~Banados, C.~Teitelboim and J.~Zanelli,
  Phys.\ Rev.\ Lett.\  {\bf 69}, 1849 (1992)
  [hep-th/9204099].







\end{thebibliography}
\end{document}